\documentclass{article}
\pdfoutput=1
\usepackage{amsmath,amssymb,graphicx,microtype,hyperref}

\usepackage{framed}
\def\be{\begin{eqnarray}}
\def\ee{\end{eqnarray}}
\def\nn{\nonumber}

\def\p{\partial}

\def\Res{{\rm Res}\,}
\def\GL{{\rm GL}\,}

\def\l[{\phantom.[}

\def\Schur{\chi}

\def\Hilb{{\rm Hilb}\,}

\hoffset= - 1.0in         

\begin{document}

\hfill ITEP/TH-17/16

\hfill IITP/TH-13/16

\bigskip

\bigskip

\centerline{\Large{
On Factorization of Generalized Macdonald Polynomials
}}

\bigskip

\bigskip

\centerline{\bf Ya.Kononov}

\bigskip

{\footnotesize
\centerline{{\it
Landau Institute for Theoretical Physics, Chernogolovka, Russia}}

\centerline{{\it
HSE, Math Department, Moscow, Russia
}}
}

\bigskip

\bigskip

\centerline{\bf A.Morozov }

\bigskip

{\footnotesize
\centerline{{\it
ITEP, Moscow 117218, Russia}}

\centerline{{\it
Institute for Information Transmission Problems, Moscow 127994, Russia
}}

\centerline{{\it
National Research Nuclear University MEPhI, Moscow 115409, Russia
}}
}

\bigskip

\bigskip

\bigskip

\bigskip

\centerline{ABSTRACT}

\bigskip

{\footnotesize
A remarkable feature of Schur functions -- the common eigenfunctions of
cut-and-join operators from $W_\infty$ -- is that they factorize
at the peculiar two-parametric topological locus in the space of time-variables,
what is known as the hook formula for quantum dimensions of representations
of $U_q(SL_N)$ and plays a big role in various applications.
This factorization  survives at the level of Macdonald polynomials.
We look for its further generalization to {\it generalized} Macdonald polynomials (GMP),
associated in the same way with the toroidal Ding-Iohara-Miki algebras,
which play the central role in modern studies in Seiberg-Witten-Nekrasov theory.
In the simplest case of the first-coproduct eigenfunctions, where GMP
depend on just two sets of time-variables, we discover a weak factorization --
on a codimension-one slice
 of the topological locus,
what is already a very non-trivial property, calling for proof and better
understanding.

}

\bigskip

\bigskip

\bigskip

\bigskip

Generalized Macdonald polynomials (GMP) \cite{GMP}
play a constantly increasing role
in modern studies of $6d$ version \cite{6dAGT,DIM}
of AGT relations \cite{AGT,AGTmamo} and spectral dualities \cite{specdu}.
At the same time they are relatively new special functions,
far from being thoroughly  understood and clearly described.
They are deformations of the generalized Jack polynomials introduced in \cite{MorSm}.
Even the simplest questions about them are yet unanswered.
In this letter we address one of them -- what happens
to the hook formulas for classical, quantum and Macdonald dimensions at the level of GMP.
We find that they survive, but only partly -- on a one-dimensional
line in the space of time-variables.
Lifting to the entire 2-dimensional topological locus remains to be found.

We begin by reminding that the Schur functions $\Schur_R\{p\}$,
depend on representation (Young diagram) $R$ and on infinitely many
time-variables $p_k$ (actually, particular $\Schur_R$ depends only on $p_k$
with $k\leq |R|=\#\ {\rm boxes\ in \ }R$).
They get nicely factorized
on peculiar two-dimensional {\it topological locus}

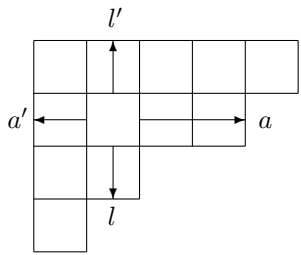
\be
p_k = p_k^*\equiv \frac{1-A^k}{1-t^k}
\label{tolo}
\ee
\be
\ \ \ \ \ \ \ \
\Schur^*_R(A,t) = \prod_{\square\in R}  t^{l'(\square)}\cdot
\prod_{\square\in R}
\frac{1-A\cdot t^{a'(\square) - l'(\square)}}
{1-t^{a(\square) + l (\square) + 1}}
\label{hookSchur}
\ee
\begin{picture}(120,0)(0,15)
\put(20,100){\line(1,0){100}}
\put(20,80){\line(1,0){100}}
\put(20,60){\line(1,0){80}}
\put(20,40){\line(1,0){40}}
\put(20,20){\line(1,0){20}}
\put(20,100){\line(0,-1){80}}
\put(40,100){\line(0,-1){80}}
\put(60,100){\line(0,-1){60}}
\put(80,100){\line(0,-1){40}}
\put(100,100){\line(0,-1){40}}
\put(120,100){\line(0,-1){20}}

\put(40,70){\vector(-1,0){20}}
\put(50,80){\vector(0,1){20}}
\put(60,70){\vector(1,0){40}}
\put(50,60){\vector(0,-1){20}}

\put(10,67){\mbox{$a'$}}
\put(48,105){\mbox{$l'$}}
\put(105,67){\mbox{$a$}}
\put(48,30){\mbox{$l$}}

\end{picture}

\noindent
Coarm $a'$ and coleg $l'$ are  the ordinary coordinates of the box in the diagram.
To keep notation continuous throughout the text, in (\ref{tolo}) we call the relevant parameter $t$, not $q$,
from the very beginning.

\newpage

It is often convenient to ignore the simple overall coefficient
and substitute the product formulas like (\ref{hookSchur}) by a polynomial
expression for {\it plethystic logarithm}
\be
\left(\prod_{\square\in R}  t^{l'(\square)}\right)^{-1}\!\!\!\!\!\cdot \Schur_R^* =
{\cal S}^\bullet\!\left(\sum_{\square \in R} t^{a(\square) + l (\square) + 1} -
A\cdot t^{a'(\square) - l'(\square)}\right)
%
\label{hookPlogSchur}
\ee
where we use the definition
\be
{\cal S}^\bullet (f) = \exp\left(\sum_k \frac{f(t_0^k,\ldots,t_n^k)}{k}\right)
\ee
for $A=t_0$ and $t=t_1$.
Plethystic exponential ${\cal S}^\bullet(f)$ is the character of the symmetric algebra of $f$
as a representation of
$\mathbb{C}^*_{t_0}\times ... \times \mathbb{C}^*_{t_n}$.
By factorization of ${\cal S}^\bullet (f)$ we actually mean  that
its plethystic  logarithm $f$ is a polynomial or, more generally,
a rational function, with integer coefficients.
The celebrated Gopakumar-Ooguri-Vafa hypothesis \cite{GOV} is
that such are not only the quantum dimensions, but also
the HOMFLY polynomials of arbitrary knots
(in this context the quantum dimensions are associated with the unknots).

Since there is a {\it double} product/sum in   (\ref{hookSchur}) and (\ref{hookPlogSchur}),
it is natural to consider their {\it refinement}, where a second $t$-parameter
is introduced, usually called $q$ (for knots this means going from HOMFLY to super- and hyper-polynomials
\cite{GSV}).
The refined version of Schur is Macdonald polynomial \cite{Mac} $M^{q,t}_R\{p\}$,
which -- unlike Schur functions -- explicitly depends on $q$ and $t$.
If $t$ is the same as in (\ref{tolo}), then (\ref{hookSchur}) lifts to
\be
M^*_R(A,t) = M^{q,t}_R\{p^*\} = \prod_{\square\in R}  t^{l'(\square)}\cdot
\prod_{\square \in R}
\frac{1-Aq^{a'(\square)} t^{-l'(\square)}}{1-q^{a(\square)} t^{l(\square)+1}}
\label{hookMac}
\ee
\be
\left(\prod_{\square\in R}  t^{l'(\square)}\right)^{-1}\!\!\!\!\!\cdot M_R^*
= {\cal S}^\bullet\!\left(
\sum_{\square \in R} q^{a(\square)} t^{l(\square)+1} - A \cdot q^{a'(\square)} t^{-l'(\square)}\right) =
{\cal S}^\bullet\Big( \mu_R^{**} - A \cdot \nu_R^{**}\Big)
\ee
We introduced  a special notation for the character of the Young diagram
\be
\nu^{**}_Y(q,t) = \sum_{\square \in Y} q^{-a'(\square)} t^{l'(\square)}
\ee
and
\be
\mu^{**}_R(q,t)
= \sum_{\square \in R} {q^{a(\square)}} {t^{l(\square)+1}}
\ee
which fully describe the plethystic logarithm of Macdonald dimension $M_R^*$.
Additional star in the label of $\mu$ and $\nu$ emphasizes that they do not depend on $A$.

The quantities $\Schur_R^*$ and $M_R^*$ are deformations of dimensions
of representation $R$ of $SL_N$-algebras, which factorize due to Weyl formulas,
and are therefore called
quantum and Macdonald dimensions.
In the former case they can be considered as graded dimension of representation $R$
of $U_t(SL_N)$, while in the latter case there is still no commonly-accepted
group-theory interpretation.
In the absence of such interpretation Macdonald polynomials are defined not as characters,
but by other less straightforward methods -- of which generalized to GMP is currently
only the definition as eigenvectors of Calogero-like Hamiltonians.
Factorization in these terms is not straightforward and is in fact a separation-of-variables
phenomenon, more-or-less equivalent to integrability of the associated theory.
Its actual {\it derivation} on these lines is usually quite tedious.
However, the very {\it fact} that factorization occurs can be observed experimentally,
far before the proofs. derivations and real understanding.
Our goal in this letter is to {\it search} for such evidence in the case of GMP.

GMP depends on set of Young diagrams and a set of time variables -- in what follows we
consider the simplest non-trivial case, when there are {\it two}: two Young diagrams and
two sets of times. Then just one more deformation parameter adds to $t$ and $q$,
we call it $Q$. Thus the GMP in question will be
denoted by ${\cal M}^{q,t,Q}_{A,B}\{p,\bar p\}$ and, in full analogy with the
Schur functions \cite{MMN},  they are eigenfunctions of the
quantum cut-and-join operator,  which is the Hamiltonian
$\Delta_{\rm DIM}(E(z))$
of the DIM algebra (see \cite{6dAGT,DIM} for details of the definition):
\be
H_1 = \frac{1}{t-1}\cdot \Res \left[
\exp\left(
\sum_{n \geq 1} \frac{1-t^{-n}}{n} z^n p_n
\right)
\exp\left(
\sum_{n \geq 1} (q^n-1) z^{-n} \frac{\p}{\p p_n}
\right)\right.
+\\
+
Q^{-1} \left.
\left(
\exp\left(
\sum_{n \geq 1} \frac{1-t^{-n}}{n} z^n ((1-t^n q^{-n}) p_n + \bar p_n)
\right)
\right)
\exp\left(
\sum_{n \geq 1} (q^n-1) z^{-n} \frac{\p}{\p p_n} - 1
\right)
\right]
\ee
Somewhat remarkably, just a single Hamiltonian is needed
to describe the whole set of GMP -- no higher Hamiltonians are needed,
because all its eigenvalues are non-degenerate.
As to label $1$ in $H_1$, it refers to the {\it first} coproduct in DIM,
higher coproducts provide Hamiltonians
for GMP,depending on more time variables.

\bigskip

Explicitly, the simplest GMP in a "natural" normalization --
which will appear consistent with the factorization property -- are:

\vspace{-1cm}

{\footnotesize
\begin{multline}
\\
M([\ ],[\ ]) =   1   \\
M([\ ],[1]) =   \bar{p}_1-\frac{p_1 (q-t)}{q (Q-1)}   \\
M([1],[\ ]) =   p_1   \\
M([\ ],[2]) =   \frac{p_1 (q+1) (t-1) \bar{p}_1 (q-t)}{(q-Q) (q t-1)}+\frac{(q+1) (t-1) \bar{p}_1^2}{2 (q t-1)}+\\
+\frac{(q-1) (t+1) \bar{p}_2}{2 (q t-1)}-\frac{p_1^2 (q+1) (t-1) (q-t) \left(q^2+q Q t-q t-Q t\right)}{2 q^2 (Q-1) (q-Q) (q t-1)}-\\
-\frac{p_2 (q-1) (t+1) (q-t) \left(q^2-q Q t+q t-Q t\right)}{2 q^2 (Q-1) (q-Q) (q t-1)}   \\
M([\ ],[1,1]) =   -\frac{p_1 \bar{p}_1 (q-t)}{q (Q t-1)}+\frac{\bar{p}_1^2}{2}-\frac{\bar{p}_2}{2}+\frac{p_1^2 (q-t) \left(q-Q t^2+Q t-t\right)}{2 q^2 (Q-1) (Q t-1)}-\\
-\frac{p_2 (q-t) \left(q-Q t^2-Q t+t\right)}{2 q^2 (Q-1) (Q t-1)}   \\
M([1],[1]) =   p_1 \bar{p}_1-\frac{p_1^2 (q-t) (q Q t+q Q-Q t+Q-2 t)}{2 q (q Q-1) (Q-t)}+\frac{p_2 (q-1) Q (t+1) (q-t)}{2 q (q Q-1) (Q-t)}   \\
M([2],[\ ]) =   \frac{p_1^2 (q+1) (t-1)}{2 (q t-1)}+\frac{p_2 (q-1) (t+1)}{2 (q t-1)}   \\
M([1,1],[\ ]) =   \frac{p_1^2}{2}-\frac{p_2}{2}  \\
\nn
\end{multline}
}

\vspace{-0.8cm}

\noindent
When all $p_k = 0$, GMP become just ordinary Macdonald polynomials of $\bar p_k$:
\be
\left.M_{Y_1 Y_2}\right|_{p_k = 0} = \delta_{Y_1, \emptyset} \cdot M_{Y_2}(\bar p_k),
\ee
and thus factorize at the topological locus (1)
of $\bar p$, i.e. at
\be
p_k=p_k^*=0,\  \  \  \  \  \  \  \bar p_k=\bar p_k^*=\frac{1-\bar A^k}{1-t^k}
\label{bartolo}
\ee
However, for $\bar p_k = 0$, they remain quite complicated functions of $p_k$:
\begin{multline}
\left.M_{[1],[1]}\right|_{\bar p_i = 0} = {\frac {p_{{2}}Q \left( -1+q \right)  \left( -t+q \right)
 \left( t+1 \right) }{ 2\left( Q-t \right)  \left( Qq-1 \right) q}}
 -  \,{\frac {{p_{{1}}}^{2} \left( -t+q \right)  \left( Qqt+Qq-Qt+Q-2\,t
 \right) }{ 2\left( Q-t \right)  \left( Qq-1 \right) q}} = \\
 = \frac{q-t}{q(1-Qq)} M_{[2]} + \frac{(1-q)(q-t)(1+t)}{(qt-1) q (Q-t)} M_{[1,1]}
\end{multline}
which do not look much simpler than the general expressions for $\bar p_k\neq 0$.
Already from these examples it is clear that GMP are non-trivial functions of $Q$.
Moreover, this complexity can seem to persist in restriction to topological locus.

However, this is not quite the case if one looks at the right quantities --
our claim is that at $A=0$ and $\bar p=0$ plethystic logarithm is just {\it linear} in $Q$
with a further factorized coefficient:
\be
\boxed{
\left(\prod_{\square \in Y_1} (-q^{a'(\square)} t)  \prod_{\square \in Y_2} (-q^{a'(\square)} t)
\right)^{-1}
\!\!\!\! \cdot M_{Y_1,Y_2}^{**}
= {\cal S}^\bullet\Big(\mu_{Y_1}^{**}+\mu_{Y_2}^{**} - t q^{-1} \nu_{Y_2}^{**}
+ Q\cdot \Upsilon_{Y_1}^{**} \cdot \nu_{Y_2}^{**}\Big)
}
\label{factconj}
\ee
see also (\ref{factformula}) below for a more explicit expression.
Here $**$ denotes the special locus
\be
p_i^{**} = \frac{1}{1-t^{-i}}, \ \ \ \bar p_i = 0,
\label{stolo}
\ee
and $\Upsilon$ is made from  the dual character $\bar \nu^{**}(q,t) = \nu^{**}(q^{-1}, t^{-1}) $:
\be
\boxed{
\Upsilon_{Y_1}^{**} = 1-(1-q)(1-t^{-1})\cdot \bar \nu_{Y_1}^{**}
}
\ee
Note that the switch $p_k\leftrightarrow \bar p_k$ is
a non-trivial operation for GMP,
described by the action of the DIM-algebra ${\cal R}$-matrix
\cite{instR}.
This implies an amusing interplay between ${\cal R}$-matrix
structure and the two very different branches of factorization locus
(\ref{bartolo}) and (14).

\bigskip

\noindent
We checked the factorization conjecture (\ref{factconj}) up to level five, i.e. for $|Y_1|+|Y_2|\leq 5$.
Here are particular examples:

\bigskip

\centerline{
\begin{footnotesize}
\begin{tabular}{|c|c|l|l|}
\hline
&&&\\
$Y_1$ & $Y_2$ & $\ \ \ \ \ \ \ \ \ {\rm Plethystic\ logarithm\ of\ } {M_{Y_1,Y_2}^{**}}$
& \ \ \ \ \ \ \ \ \ \ \ \ \ \ \ \ $M_{Y_1,Y_2}^{**}$ \\
&&&\\
\hline
[\ ] & [1] & $-\frac{t}{q}+Q+t$ & $-\frac{t (q-t)}{q (Q-1) (t-1)}$ \\
\hline
[\ ] & [2] & $-\frac{t}{q^2}+\frac{Q}{q}+q t-\frac{t}{q}+Q+t$ & $ -\frac{t^2 (q-t) \left(q^2-t\right)}{q (Q-1) (t-1) (q-Q) (q t-1)}$ \\
\hline
[\ ] & [1,1] & $-\frac{t^2}{q}-\frac{t}{q}+Q t+Q+t^2+t$ & $\frac{t^2 (q-t) \left(q-t^2\right)}{q^2 (Q-1) (t-1)^2 (t+1) (Q t-1)}$ \\
\hline
[1] & [1] &  $   -\frac{q Q}{t}+q Q-\frac{t}{q}+\frac{Q}{t}+2 t  $ & $-\frac{t^2 (q-t) (q Q-t)}{q (t-1)^2 (q Q-1) (Q-t)}$\\
\hline
[\ ] & [3] & $ -\frac{t}{q^3}+\frac{Q}{q^2}+q^2 t-\frac{t}{q^2}+\frac{Q}{q}+q t-\frac{t}{q}+Q+t   $ & $-\frac{t^3 (q-t) \left(q^2-t\right) \left(q^3-t\right)}{(Q-1) (t-1) (q-Q) \left(q^2-Q\right) (q t-1) \left(q^2 t-1\right)}$\\
\hline
[\ ] & [2,1] & $ -\frac{t}{q^2}+\frac{Q}{q}+q t^2-\frac{t^2}{q}-\frac{t}{q}+Q t+Q+2 t    $ & $\frac{t^3 (q-t) \left(q^2-t\right) \left(q-t^2\right)}{q^2 (Q-1) (t-1)^2 (q-Q) \left(q t^2-1\right) (Q t-1)}$\\
\hline
[\ ] & [1,1,1] &  $  -\frac{t^3}{q}-\frac{t^2}{q}-\frac{t}{q}+Q t^2+Q t+Q+t^3+t^2+t  $ & $-\frac{t^3 (q-t) \left(q-t^2\right) \left(q-t^3\right)}{q^3 (Q-1) (t-1)^3 (t+1) \left(t^2+t+1\right) (Q t-1) \left(Q
   t^2-1\right)}$ \\
\hline
[1] & [2] & $  -\frac{t}{q^2}-\frac{q Q}{t}+\frac{Q}{q t}+q Q+q t-\frac{t}{q}+Q+2 t   $ & $-\frac{t^3 (q-t) \left(q^2-t\right) (q Q-t)}{q (Q-1) (t-1)^2 (q Q-1) (q t-1) (q t-Q)}$ \\
\hline
[1] & [1,1] & $  q Q t-\frac{q Q}{t}-\frac{t^2}{q}-\frac{t}{q}+\frac{Q}{t}+Q+t^2+2 t   $ & $\frac{t^3 (q-t) \left(q-t^2\right) (q Q-t)}{q^2 (Q-1) (t-1)^3 (t+1) (Q-t) (q Q t-1)}$\\
\hline
[2] & [1] & $   -\frac{q^2 Q}{t}+q^2 Q+q t-\frac{t}{q}+\frac{Q}{t}+2 t  $ & $-\frac{t^3 (q-t) \left(q^2 Q-t\right)}{(t-1)^2 \left(q^2 Q-1\right) (q t-1) (Q-t)}$\\
\hline
[1,1] & [1] & $  -\frac{q Q}{t^2}+q Q-\frac{t}{q}+\frac{Q}{t^2}+t^2+2 t   $ & $-\frac{t^3 (q-t) \left(q Q-t^2\right)}{q (t-1)^3 (t+1) (q Q-1) \left(Q-t^2\right)}$\\
\hline
[3] & [1] & $  -\frac{q^3 Q}{t}+q^3 Q+q^2 t+q t-\frac{t}{q}+\frac{Q}{t}+2 t   $ & $-\frac{q^2 t^4 (q-t) \left(q^3 Q-t\right)}{(t-1)^2 \left(q^3 Q-1\right) (q t-1) \left(q^2 t-1\right) (Q-t)}$\\
\hline
[2,1]& [1] & $ -\frac{q^2 Q}{t}+q^2 Q-\frac{q Q}{t^2}+\frac{q Q}{t}+q t^2-\frac{t}{q}+\frac{Q}{t^2}+3 t    $ & $-\frac{t^4 (q-t) \left(q^2 Q-t\right) \left(q Q-t^2\right)}{(t-1)^3 \left(q^2 Q-1\right) \left(q t^2-1\right)
   \left(Q-t^2\right) (q Q-t)}$ \\
\hline
[1,1,1] & [1] & $  -\frac{q Q}{t^3}+q Q-\frac{t}{q}+\frac{Q}{t^3}+t^3+t^2+2 t   $ & $-\frac{t^4 (q-t) \left(q Q-t^3\right)}{q (t-1)^4 (t+1) \left(t^2+t+1\right) (q Q-1) \left(Q-t^3\right)}$ \\
\hline
[2] & [2] & $ -\frac{q^2 Q}{t}+q^2 Q-\frac{t}{q^2}-\frac{q Q}{t}+\frac{Q}{q t}+q Q+2 q t-\frac{t}{q}+\frac{Q}{t}+2 t    $ & $-\frac{t^4 (q-t) \left(q^2-t\right) (q Q-t) \left(q^2 Q-t\right)}{(t-1)^2 (q Q-1) \left(q^2 Q-1\right) (q t-1)^2 (Q-t) (q
   t-Q)}$ \\
\hline
[2] & [1,1] & $  q^2 Q t-\frac{q^2 Q}{t}-\frac{t^2}{q}+q t-\frac{t}{q}+\frac{Q}{t}+Q+t^2+2 t   $ & $\frac{t^4 (q-t) \left(q-t^2\right) \left(q^2 Q-t\right)}{q (Q-1) (t-1)^3 (t+1) (q t-1) (Q-t) \left(q^2 Q t-1\right)}$ \\
\hline
[1,1] & [2] & $  -\frac{t}{q^2}-\frac{q Q}{t^2}+\frac{Q}{q t^2}+q Q+q t-\frac{t}{q}+Q+t^2+2 t   $ & $-\frac{t^4 (q-t) \left(q^2-t\right) \left(q Q-t^2\right)}{q (Q-1) (t-1)^3 (t+1) (q Q-1) (q t-1) \left(q t^2-Q\right)}$ \\
\hline
[1,1] & [1,1] & $   -\frac{q Q}{t^2}+q Q t-\frac{q Q}{t}+q Q-\frac{t^2}{q}-\frac{t}{q}+\frac{Q}{t^2}+\frac{Q}{t}+2 t^2+2 t  $ & $\frac{t^4 (q-t) \left(q-t^2\right) (q Q-t) \left(q Q-t^2\right)}{q^2 (t-1)^4 (t+1)^2 (q Q-1) (Q-t) \left(Q-t^2\right) (q Q
   t-1)}$ \\
\hline
[1] & [3] & $  -\frac{t}{q^3}+\frac{Q}{q^2 t}+q^2 t-\frac{t}{q^2}-\frac{q Q}{t}+q Q+\frac{Q}{q}+q t-\frac{t}{q}+Q+2 t   $ & $-\frac{t^4 (q-t) \left(q^2-t\right) \left(q^3-t\right) (q Q-t)}{(Q-1) (t-1)^2 (q-Q) (q Q-1) (q t-1) \left(q^2 t-1\right)
   \left(q^2 t-Q\right)}$ \\
\hline
[1] & [2,1] & $  -\frac{t}{q^2}+q Q t-\frac{q Q}{t}+\frac{Q}{q t}+q t^2-\frac{t^2}{q}-\frac{t}{q}+2 Q+3 t   $ & $\frac{t^4 (q-t) \left(q^2-t\right) \left(q-t^2\right) (q Q-t)}{q^2 (Q-1)^2 (t-1)^3 \left(q t^2-1\right) (q t-Q) (q Q t-1)}$ \\
\hline
[1] & [1,1,1] & $  q Q t^2-\frac{q Q}{t}-\frac{t^3}{q}-\frac{t^2}{q}-\frac{t}{q}+Q t+\frac{Q}{t}+Q+t^3+t^2+2 t   $ & $-\frac{t^4 (q-t) \left(q-t^2\right) \left(q-t^3\right) (q Q-t)}{q^3 (Q-1) (t-1)^4 (t+1) \left(t^2+t+1\right) (Q-t) (Q t-1)
   \left(q Q t^2-1\right)}$ \\
\hline
[2,1,1] & [1] & $  -\frac{q^2 Q}{t}+q^2 Q-\frac{q Q}{t^3}+\frac{q Q}{t}+q t^3-\frac{t}{q}+\frac{Q}{t^3}+t^2+3 t   $ & $-\frac{t^5 (q-t) \left(q^2 Q-t\right) \left(q Q-t^3\right)}{(t-1)^4 (t+1) \left(q^2 Q-1\right) \left(q t^3-1\right)
   \left(Q-t^3\right) (q Q-t)}$ \\
\hline
[2,2] & [1] & $  -\frac{q^2 Q}{t^2}+q^2 Q+q t^2+q t-\frac{t}{q}+\frac{Q}{t^2}+t^2+2 t   $ & $-\frac{q t^5 (q-t) \left(q^2 Q-t^2\right)}{(t-1)^3 (t+1) \left(q^2 Q-1\right) (q t-1) \left(q t^2-1\right)
   \left(Q-t^2\right)}$ \\
\hline
\end{tabular}
\end{footnotesize}
}

\bigskip

\noindent
$M_{Y_1,[\ ]}=M_{Y_1}$ are just the ordinary Macdonald polynomials, factorized
according to (\ref{hookMac}),  and they are omitted from the table.

\bigskip

Despite (\ref{stolo}) is a relatively small slice of what one could expect
from refinement of the topological locus,
the property (\ref{factconj}) looks quite spectacular and mysterious.
Its understanding can provide new insights about both the GMP
and integrality conjectures.
GMP are clearly simpler and closer to conventional group theory than
knot (super)polynomials -- and (\ref{factconj}) is naturally  more structured
than one can expect for generic knots,
however, its continuation to $A\neq 0$ can already be of more general type.

\bigskip

Factorization does not survive neither the deformation to non-vanishing
$\bar p_i^*=\frac{1-\bar A^i}{1-\bar t^{-i}}$ for any $\bar t$,
nor the   $A$-deformation to $p_i = \frac{1-A^i}{1-t^{-i}}, \; \bar p_i = 0$:
\begin{multline}
M^*([1],[2]) = (A-1) t^3 (q-t) \left(A^2 q^3 Q^2 t^2-A^2 q^3 Q t^2+A^2 q^3 Q t
-A^2 q^2 Q^2 t^2+A^2 q^2 Q^2 t-A^2 q^2 Q t^2+A^2 q^2   t^2-\right.\\
  \left.-A^2 q^2 t-A^2 q Q^2 t^2+A^2 q Q t^2-2 A^2 q Q t+A^2 q Q+A^2 Q^2 t^2-A^2 Q^2 t
  +A^2 Q t^2-A q^4 Q t-A q^3 Q^2 t+A q^3 Q   t^2-\right.\\
   \left. - A q^3 Q+A q^3 t+A q^2 Q t^2+2 A q^2 Q t-A q^2 Q-A q^2 t^2+A q^2 t+A q Q^2 t
   -A q Q t^2+A q Q t-A q t^2-A Q t^2+\right. \\
   \left. +q^4 Q-q^3
   t-q^2 Q t+q t^2\right)/({q^2 (Q-1) (t-1)^2 (q Q-1) (q t-1) (q t-Q)})
\end{multline}
while at $A=0$ we return to a nicely factorized
\be
M^{**} ([1],[2]) = -\frac{t^3 (q-t) \left(q^2-t\right) (q Q-t)}{q (Q-1) (t-1)^2 (q Q-1) (q t-1) (q t-Q)}
\ee
As an $A$-series, plethystic logarithm of the ratio $M^*_{[1],[1]}/M^{**}_{[1],[1]}$ is:
\begin{multline}
-\frac{q Q t+q Q-Q t+Q-2 t}{q Q-t} A  + \frac{(q-1) Q (t-1) t (q Q-1) (Q-t)}{(q Q-t)^2 (q Q+t)} A^2 + \\
+ \frac{(q-1) Q (t-1) t (q
   Q-1) (Q-t) \left(q^2 Q^2 t+q Q^2-Q t^2-t^2\right)}{(q Q-t)^3 \left(q^2 Q^2+q Q t+t^2\right)} A^3+...
\end{multline}
The first term at the r.h.s. resembles the $Q$-linear term in (\ref{factconj}).
An additional surprise is that the second term also factorizes, and really bad things
happen only in the order $A^3$.

\bigskip

Along with factorization, Schur functions satisfy the  Cauchy identity:
\be
\sum_R \Schur_R\{p\}\Schur_R\{\bar p\} = \exp \left(\sum_{k=1}^\infty \frac{p_n\bar p_n}{n}\right)
\label{Cid}
\ee
Already in the simplest case of Schur {\it polynomials} (where only single-line Young diagrams $R=[n]$,
i.e. pure symmetric representations, contribute), combination of this identity and
factorization provide a remarkable product formula, identifying the $t$-exponent
and Pochhammer symbol:
\be
\sum_{n=0} \frac{z^n}{[n]!} =
\sum_{n \geq 0} \frac{z^n}{\prod_{1 \leq i \leq n} (1-t^i)} =
\prod_{n\geq 0} \frac{1}{1-z t^n}
\label{qexpPol}
\ee
This can be also considered as a
formula for the Euler characteristic of the structure sheaf of $\Hilb(\mathbb{C},n)$.
In Miwa coordinates $p_k=\sum_i x_i^k$ Cauchy identity (\ref{Cid})
can be considered as following from the decomposition
$
{\cal S}^\bullet (V\otimes W) = \oplus_{R} \Schur_R V \otimes \Schur_R W
$
as $\GL(V) \times \GL(W)$-modules:
\be
\sum_R \Schur_R[x] \Schur_R [y] = \prod_{i,j} \frac{1}{1-x_i y_j}
\ee
If we restrict both  $p_i$ and $\bar p_i$ to the topological locus,
\be
p_n^*=\frac{1-A^n}{1-t^{-n}}, \ \ \ \ \ \ \ \ \bar p_n^*=\frac{1-B^n}{1-q^{-n}}
\ee
then
\be
 \sum_{R} z^{|R|} \prod_{\square \in R} \frac{t^{l'(\square)}
- A t^{a'(\square)}}{1-t^{a(\square)+l(\square)+1}}\cdot
\frac{q^{l'(\square)} - B q^{a'(\square)}}{1-q^{a(\square)+l(\square)+1}}
= \exp \left(\sum_{n=1}^{\infty}
\frac{ z^n}{n} \frac{1-A^n}{1-t^n} \frac{1-B^n}{1-q^n}
\right)
\ee

For Macdonald polynomials it deserves taking the proper form of the Cauchy identity:
\be
\sum_n z^n \sum_{|R|=n} M_R^{q,t} M_{\bar R}^{t^{-1}\!,\,q^{-1}} = \exp \left(
\sum_{n \geq 1} - \frac{(-z)^n}{n} p_n \bar p_n
\right).
\ee
It is related to a more complicated but better known version
\be
\sum_n z^n \sum_{|R|=n} \left(
\prod_{\square \in R} \frac{1-q^{a(\square)+1} t^{l(\square)}}{1 - q^{a(\square)} t^{l(\square)+1}}
\right)^{-1}
M_R (p) M_R (\bar p) = \exp \left(
\sum_{k \geq 1} \frac{1-t^k}{1-q^k} \frac{p_k \bar p_k}{k}
\right)
\ee
through the transposition identity \cite{transp,GMP}
\be
M_R^{q,t} \left( -\frac{1-q^i}{1-t^i} p_i \right) = \prod_{\square \in R} \left(
- \frac{1-q^{a(\square)+1} t^{l(\square)}}{1-q^{a(\square)} t^{l(\square)+1}}
\right) M_{R'}^{t,q} (p_i)
\ee
It actually simplifies greatly at the topological locus:
\be
\sum_R z^{|R|} \prod_{\square \in R}
\left(
t^{2 l'(\square)} \frac{(1-A q^{a'(\square)} t^{-l'(\square)})(1-B q^{a'(\square)}
t^{-l'(\square)})}{(1-q^{a(\square)} t^{l(\square)+1}) (1-q^{a(\square)+1} t^{l(\square)}}
\right) = \exp\left(
\sum_{k \geq 1} \frac{z^n}{n} \frac{(1-A^n)(1-B^n)}{(1-t^n)(1-q^n)}
\right)
\label{taut}
\ee
This formula can be seen as equivariant Euler characteristic of a certain tautological bundle
over the Hilbert scheme.
Refs.\cite{CNOP} provide factorization formulas for the equivariant Euler characteristic
of the sheaf of differential forms:
\be
\sum_{R} z^{|R|} \prod_{\square \in R} \frac{(1-M t_1^{-a(\square))} t_2^{l(\square)+1})
(1-M t_1^{a(\square)+1} t_2^{-l(\square)})}{(1-t_1^{-a(\square)} t_2^{l(\square)+1})
(1-t_1^{a(\square)+1} t_2^{-l(\square)})}
=
{\cal S}^\bullet \!\left( \frac{z}{1-M z} \frac{(1-M t_1)(1-M t_2)}{(1-t_1)(1-t_2) }\right)
\label{diff}
\ee
In the case $A = B = 0$ and $M = 0$ both (\ref{taut}) and (\ref{diff})
represent the equivariant Euler characteristic of the structure sheaf, and the formulas become identical.

For GMP the Cauchy identity is    \cite{GMP}
\be
\sum_{Y_1, Y_2} z^{|Y_1|+|Y_2|} M_{Y_1, Y_2}^{q,t} M_{Y_2^t, Y_1^t}^{t^{-1}\!,\,q^{-1}} = \exp \sum_{n \geq 1}
\left[
\frac{(-z)^n}{n} \left(
\left(
1-\frac{t^n}{q^n}
\right) p_n q_n + \bar p_n q_n + p_n \bar q_n
\right)
\right]
\ee
and our factorization conjecture (\ref{factconj}) implies the following analogue of (\ref{qexpPol}):
\be
\boxed{
\sum_{Y_1, Y_2} z^{|Y_1|+|Y_2|} M_{Y_1, Y_2}^{q,t\,**} M_{Y_2^t, Y_1^t}^{t^{-1}\!,\, q^{-1}\,**} =
\exp \left[
\sum_{n \geq 1} \frac{(-z)^n}{n} \left(
 \frac{1-t^n/q^n}{(1-t^{-n})(1-q^n)}
\right)
\right]
}
\label{id}
\ee
where
\be
\boxed{
M_{Y_1, Y_2}^{q,t\,**} = \prod_{Y_1} \frac{-q^{a'} t}{1-q^{a} t^{l+1}} \prod_{Y_2}
\frac{(-q^{a'} t)(1-q^{-a'-1} t^{l'+1})}{(1-q^a t^{l+1})(1-Q q^{-a'} t^{l'})} \prod_{Y_1, Y_2}
\frac{(1-q Q q^{a_1'-a_2'} t^{l_2'-l_1'})(1-t^{-1} Q q^{a_1'-a_2'}
t^{l_2'-l_1'})}{(1-Q q^{a_1'-a_2'} t^{l_2'-l_1'}) (1-q t^{-1} Q q^{a_1'-a_2'} t^{l_2'-l_1'})}
}
\label{factformula}
\ee
and


\centerline{
$
 M_{Y_1^t, Y_2^t}^{t^{-1}, q^{-1}} (p_k, \bar p_k) =    (-1)^{|Y_1|+|Y_2|} \left( \prod_{\square \in Y_1}
 \frac{1-q^{a(\square)+1} t^{l(\square)}}{1 - q^{a(\square)} t^{l(\square)+1}}
 \prod_{\square \in Y_2}  \frac{1-q^{a(\square)+1} t^{l(\square)}}{1 - q^{a(\square)}
 t^{l(\square)+1}} \right)^{-1}\!\!\!\!\!
\cdot M_{Y_1, Y_2}^{q,t} \left(
 - \frac{1-q^k}{1-t^k} p_k,  - \frac{1-q^k}{1-t^k} \bar p_k
\right)
$
}

\bigskip

\noindent
Eq.(\ref{id}) is a {\it new} identity,
which is a generalization of (\ref{qexpPol}) with 2 additional parameters.
This formula expresses the Euler characteristic of a certain tautological bundle
on the moduli space of framed sheaves on $\mathbb{P}^2$.
The left hand side is the sum of localization contribution of the fixed points
(which are parameterized by pairs of Young diagrams) due to the K-theoretic Lefschetz fixed point formula.

\bigskip

To conclude, we investigated the possibility that {\it generalized} Macdonald polynomials
satisfy some kind of a hook factorization formula on some kind of a topological locus --
and discovered that this is indeed the case.
The main observation (conjecture) is a rather beautiful (\ref{factconj}),
where $Q$ enters only linearly -- like $A$ in (\ref{hookPlogSchur}), and the coefficient in front of it
is further decomposes into $Y_1$- and $Y_2$-dependent factors.
This result is twice surprising:
it is remarkable that factorization at all persists for GMP,
but, if it does, why only on the restricted one-parametric subspace (\ref{stolo})
and not on the naive three-parametric extension of (\ref{tolo}),
\be
p_i = \frac{1-A^i}{1-t^{-i}}, \ \ \ \ \bar p_i = \frac{1-\bar A^i}{1- t^{-i}}
\label{tolobig}
\ee
Hopefully, this small discovery will cause interest to the problem and this will help
to clarify the structure of Weyl formulas for DIM and double Hekke algebras,
which stand behind the generalized Macdonald polynomials and their properties.
The first step in this direction is already made in \cite{Znew}:
as one could anticipate from (\ref{bartolo}),
when $\bar p$ is non-vanishing,
the factorization locus increases, to
\be
p_i = -\frac{1-A^i}{1-t^{-i}}, \ \  \bar p_i = \frac{1-(t/q)^{i}}{1- t^{-i}},
\ee
but still remains of codimension one w.r.t. (\ref{tolobig}).
Despite the sign difference with (\ref{stolo}), which makes the loci, big and small,
not even intersecting, restrictions of generalized  Macdonalds are nearly the same:
\be
M^*_{Y_1,Y_2} \ \sim \  M^{**}_{Y_1,Y_2} \cdot \prod_{\square \in Y_1}
(At^{-l(\square)}-q^{-a(\square)})
\prod_{\square \in Y_2}
(At^{-l(\square)}-Q q^{-a(\square)})
\ee
Proportionality factor is a power of $t$.

The next immediate questions concern extension to GMP, depending on many
time variables (eigenfunctions of the higher coproducts of DIM),
and extension to more general representations of DIM, labeled by 3d (plane) partitions.

\section*{Acknowledgements}

We are grateful to B.Feigin, A.Mironov, Sh.Shakirov and Y.Zenkevich for helpful discussions.

\noindent
Our work is partly supported by grants
RFBR grants 16-01-00291 (Y.K.), 16-02-01021 (A.M.) by young scientist
grants 16-31-00484 (Y.K.), 15-31-20832-mol-a-ved (A.M.), by Simons Foundation (Y.K.)
and by the joint grants
15-51-52031-{HHC},
15-52-50041-YaF,
16-51-53034-GFEN,
16-51-45029-Ind


\begin{thebibliography}{12}

\bibitem{GMP}
Y.~Ohkubo, arXiv:1404.5401\\
Y.~Zenkevich
JHEP 1505 (2015) 131, arXiv:1412.8592\\
 A.~Morozov and Y.~Zenkevich,
JHEP 1602 (2016) 098, arXiv:1510.01896



\bibitem{6dAGT}
H.~Awata and H.~Kanno, JHEP 0505 (2005) 039, hep-th/0502061; IJMPA24 (2009) 2253,
arXiv:0805.0191; arXiv:0903.5383\\
M.~Taki, JHEP 0803 (2008) 048, arXiv:0710.1776 \\
H.Awata, H.Fuji, H.Kanno, M.Manabe, Y.Yamada, Adv.Theor.Math.Phys. 16 (2012) 725,
arXiv:1008.0574 \\
H. Nakajima and K. Yoshioka, math/0306198; math/0505553; math/0311058\\
A.~Mironov, A.~Morozov, Y.~Zenkevich,
 Phys.Lett. B756 (2016) 208-211, arXiv:1512.06701\\
H.Awata, H.Kanno, T.Matsumoto, A.Mironov,
    A.Morozov, An.Morozov, Y.Ohkubo and Y.Zenkevich,
JHEP 07 (2016) 1-67,    arXiv:1604.08366   \\
J.-E.~Bourgine, M.~Fukuda, Y.~Matsuo, H.~Zhang, R.-D.~Zhu,
arXiv:1606.08020

\bibitem{DIM}
J. Ding, K. Iohara, Lett. Math. Phys. 41 (1997), q-alg/9608002 \\
K. Miki, J. Math. Phys. 48 (2007) 123520 \\
V. Ginzburg, M. Kapranov and E. Vasserot, Mathem. Research Letters, 2 (1995) 147-160, q-alg/9502013 \\
H. Nakajima, Ann. of Math. (2) 160 (2004) 10571097; math/0204184; math/0204185 \\
 H. Awata, B. Feigin, A. Hoshino, M. Kanai, J. Shiraishi and S. Yanagida,
 arXiv:1106.4088 \\
S. Kanno, Y. Matsuo and S. Shiba, Phys. Rev. D84 (2011) 026007, arXiv:1105.1667 \\
S. Kanno, Y. Matsuo and H. Zhang, arXiv:1207.5658; arXiv:1306.1523 \\
N. Nekrasov and V. Pestun, arXiv:1211.2240 \\
N. Nekrasov, S. Shatashvili and V. Pestun, arXiv:1312.6689 \\
J.-E. Bourgine, Y. Matsuo and H. Zhang, arXiv:1512.02492  \\
 N. Nekrasov, arXiv:1512.05388 \\
 T. Kimura and V. Pestun, arXiv:1512.08533 \\
A. Mironov, A. Morozov, Y. Zenkevich, arXiv:1603.05467



\bibitem{AGT}
L. Alday, D. Gaiotto and Y. Tachikawa, Lett. Math. Phys. 91 (2010) arXiv:0906.3219\\
N. Wyllard, JHEP 0911 (2009) 002, arXiv:0907.2189 \\
A. Mironov and A. Morozov, Nucl. Phys. B825 (2009) 1-37, arXiv:0908.2569,
 arXiv:1012.3137

\bibitem{AGTmamo}
A.Mironov, A.Morozov, Sh.Shakirov,
JHEP 1002, 030 (2010) arXiv:0911.5721;
IJMPA25 (2010) 3173-3207, arXiv:1001.0563;
IJMPA27 (2012) 1230001,  arXiv:1011.5629;
JHEP 1102:067,2011,  arXiv:1012.3137  \\
H.Itoyama and T.Oota, arXiv:1003.2929\\
A.Mironov, A.Morozov, Sh.Shakirov, A.Smirnov,
 Nucl.Phys. B 855 (2012)  128-151, arXiv:1105.0948 \\
Y.~Zenkevich,
arXiv:1507.00519


\bibitem{specdu}
M. R. Adams, J. Harnad, J. Hurtubise, Lett. Math. Phys., Vol. 20, Num. 4, 299-308 (1990) \\
J. Harnad, Commun. Math. Phys., Vol. 166, Num. 2, 337-365 (1994), arXiv:hep-th/9301076 \\
G. Wilson, J. Reine Angew. Math. 442, 177?204 (1993)\\
M. Bertola, B. Eynard, J. Harnad, Commun. Math. Phys., Vol. 229, Num. 1, 73?120 (2002), nlin/0108049\\
V. Tarasov, A. Varchenko, Acta Applic. Mathem. 73 (2002) no. 1-2, 141?154, arXiv:math/0112005\\
 E. Mukhin, V. Tarasov, A. Varchenko, math/0610799, math/0510364, math/0605172\\
A. Mironov, A. Morozov, Y. Zenkevich, A. Zotov,
Pis'ma v ZhETF 97 (2013) 49-56, arXiv:1204.0913  \\
A.Mironov, A.Morozov, B.Runov, Y.Zenkevich and A.Zotov,
  Lett.Math.Phys. 103 (2013) 299-329, arXiv:1206.6349;
JHEP 1312 (2013) 034, arXiv:1307.1502 \\
M. Aganagic, N. Haouzi, C. Kozcaz and S. Shakirov, arXiv:1309.1687\\
M. Aganagic, N. Haouzi and S. Shakirov, arXiv:1403.3657\\
M. Taki, arXiv:1310.7509; arXiv:1401.7200 \\
V. Mitev, E. Pomoni, M. Taki and F. Yagi, JHEP 04 (2015) 052, arXiv:1411.2450 \\
S.-S. Kim, M. Taki and F. Yagi, Prog. Theor. Exp. Phys. (2015) 083B02, arXiv:1504.03672 \\
H. Hayashi, S.-S. Kim, K. Lee, M. Taki and F. Yagi, JHEP 1508 (2015) 097, arXiv:1505.04439 \\
M. Aganagic and N. Haouzi, arXiv:1506.04183\\
A. Mironov, A. Morozov, Y. Zenkevich,
JHEP 05 (2016) 1-44, arXiv:1603.00304
\bibitem{MorSm}
A.~Morozov and A.~Smirnov,
Lett.Math.Phys. 104  (2014)  585-612, arXiv:1307.2576 \\
S. Mironov, An. Morozov, Y. Zenkevich, JETP Lett. 99 (2014) 109, arXiv:1312.5732











\bibitem{GOV}
R.Gopakumar and C.Vafa,
Adv.Theor.Math.Phys. 3 (1999) 1415, hep-th/9811131 \\
H.Ooguri and C.Vafa, Nucl.Phys.B577 (2000) 419-438,  hep-th/9912123\\
M.Marino and C.Vafa, hep-th/0109064 \\
J.M.F.Labastida, M.Marino,
J.Knot theory Ramif. 11 (2002) 173, hep-th/0004196;
Comm.Math.Phys. 217 (2001) 423, hep-th/0004196 \\
J.M.F.Labastida, M.Marino and C.Vafa,
JHEP0011 (2000) 007, hep-th/0010102   \\
K.Liu and P.Peng, arXiv:0704.1526 \\
A.Mironov, A.Morozov, An.Morozov, P.Ramadevi, V.K.Singh, A.Sleptsov,
{\it New integrality tests for knot/link polynomials}

\bibitem{GSV}
S.Gukov, A.Schwarz and C.Vafa,  Lett.Math.Phys. 74 (2005) 53-74, hep-th/0412243 \\
N.Dunfield, S.Gukov and J.Rasmussen, Experimental Math. 15 (2006) 129-159, math/0505662 \\
M.Aganagic and Sh.Shakirov,
arXiv:1105.5117,
 arXiv:1202.2489,
arXiv:1210.2733  \\
P. Dunin-Barkowski, A. Mironov, A. Morozov, A. Sleptsov, A. Smirnov,
JHEP 03 (2013) 021, arXiv:1106.4305
I.Cherednik, I.Cherednik, arXiv:1111.6195; \\
E.Gorsky, A.Oblomkov, J.Rasmussen and V.Shende,  Duke Math.J. 163  (2014) 2709-2794,
 arXiv:1207.4523  \\
A. Mironov, A. Morozov, Sh. Shakirov, A. Sleptsov, arXiv:1201.3339 \\
E.Gorsky, S.Gukov and M.Stosic, arXiv:1304.3481 \\
S.Nawata, P. Ramadevi and Zodinmawia,
JHEP 1401 (2014) 126, arXiv:1310.2240 \\
S.Arthamonov and Sh.Shakirov,
arXiv:1504.02620 \\
S.Nawata and A.Oblomkov,
arXiv:1510.01795


\bibitem{Mac}
I.G.~Macdonald
{\it Symmetric Functions and Hall Polynomials},
{\em Oxford University Press}

\bibitem{MMN}
A.Mironov, A.Morozov and S.Natanzon,
Theor.Math.Phys.166 (2011) 1-22, arXiv:0904.4227

\bibitem{instR}
A.Smirnov,  arXiv:1302.0799 \\
H.Awata, H.Kanno,  A.Mironov, A.Morozov, An.Morozov, Y.Ohkubo and Y.Zenkevich,  arXiv:1608.05351

\bibitem{transp}
A.Mironov, A.Morozov and Sh.Shakirov,
J.Phys. A45 (2012) 355202,  arXiv:1203.0667

\bibitem{CNOP}
A.~Okounkov and R.~Pandharipande,
math/0411210 \\
E.~Carlsson, N.~Nekrasov and A.~Okounkov,
arXiv:1308.2465 \\
E.~Carlsson and A.~Okounkov,
arXiv:0801.2565 \\
A.~Okounkov,
arXiv:1512.07363

\bibitem{Znew} Y.Zenkevich, {\it to appear}


\end{thebibliography}
\end{document}